\documentstyle[12pt,psfig]{article}
\let\section=\subsection     \let\subsection=\subsubsection                

\begin{document}
\begin{center}
   {\large \bf Phase Transitions in Neutron Stars}\\[5mm]
   Henning Heiselberg and Morten Hjorth-Jensen \\[5mm]
   {\small \it  NORDITA \\
   Blegdamsvej 17, DK-2100 Copenhagen \O, Denmark \\[8mm] }
\end{center}

\begin{abstract}\noindent
Phase transitions in neutron stars due to formation of quark matter,
kaon condensates, etc. are discussed with particular attention to the
order of these transitions. Observational consequences of phase
transitions in pulsar angular velocities are examined.
\end{abstract}

\section{Introduction}
The physical state of matter in the interiors of neutron stars at
densities above a few times normal nuclear matter densities is
essentially unknown. Interesting phase transitions in nuclear matter
to quark matter \cite{Glendenning,HPS}, kaon \cite{Kaplan,kaon} or pion
condensates \cite{pion,vijay}, neutron and proton superfluidity
\cite{oeystein}, hyperonic matter, crystalline nuclear matter
\cite{pion}, magnetized matter, etc., have been considered.  We
discuss how these phase transitions may exist in a mixed phase, the
structures formed and in particular the order of the
transition. Observational consequences are discussed.

\section{Mixed Phases and Order of Transitions}
The mixed phase in the inner crust of neutron stars consists of
nuclear matter and a neutron gas in $\beta$-equilibrium with a
background of electrons such that the matter is overall electrically
neutral \cite{LPR}. Likewise, quark and nuclear matter can have a
mixed phase \cite{Glendenning} and possible also nuclear matter with
and without condensate of any negatively charged particles such as
$K^-$ \cite{Schaffner}, $\pi^-$, $\Sigma^-$, etc.  The quarks are
confined in droplet, rod- and plate-like structures \cite{HPS}
analogous to the nuclear matter and neutron gas structures in the
inner crust of neutron stars \cite{LPR}.  Depending on the equation of
state, normal nuclear matter exists only at moderate densities,
$\rho\sim 1-2\rho_0$.  With increasing density, droplets of quark
matter form in nuclear matter and may merge into rod- and later
plate-like structures.  At even higher densities the structures invert
forming plates, rods and droplets of nuclear matter in quark matter.
Finally pure quark matter is formed at very high densities unless the
star already has exceeded its maximum mass.

A necessary condition for forming these structures and the mixed phase
is that the additional surface and Coulomb energies of these
structures are sufficiently small. Excluding them makea the mixed
phase energetically favored \cite{Glendenning}. That is also the
case when surface energies are small (see \cite{HPS} for a quantitative
condition). If they are
too large the neutron star will 
have a core of pure quark matter with a mantle of nuclear matter
surrounding and the two phases are coexisting by an ordinary first
order phase transition.

The quark and nuclear matter mixed phase has continuous pressures and
densities \cite{Glendenning} when surface and Coulomb energies are
excluded.  There are at most two second order phase transitions. Namely,
at a lower density, where quark matter first appears in nuclear
matter, and at a very high density, where all nucleons are finally
dissolved into quark matter, if the star is gravitationally stable at
such high central densities.  However, due to the finite Coulomb and
surface energies associated with forming these structures, the
transitions change from second to first order at each topological change in
structure \cite{HPS}. If the surface and Coulomb energies are very
small the transitions will be only weakly first order but there may be
several of them.

\noindent
\begin{minipage}[b]{15cm}
{\centering
\mbox{\psfig{file=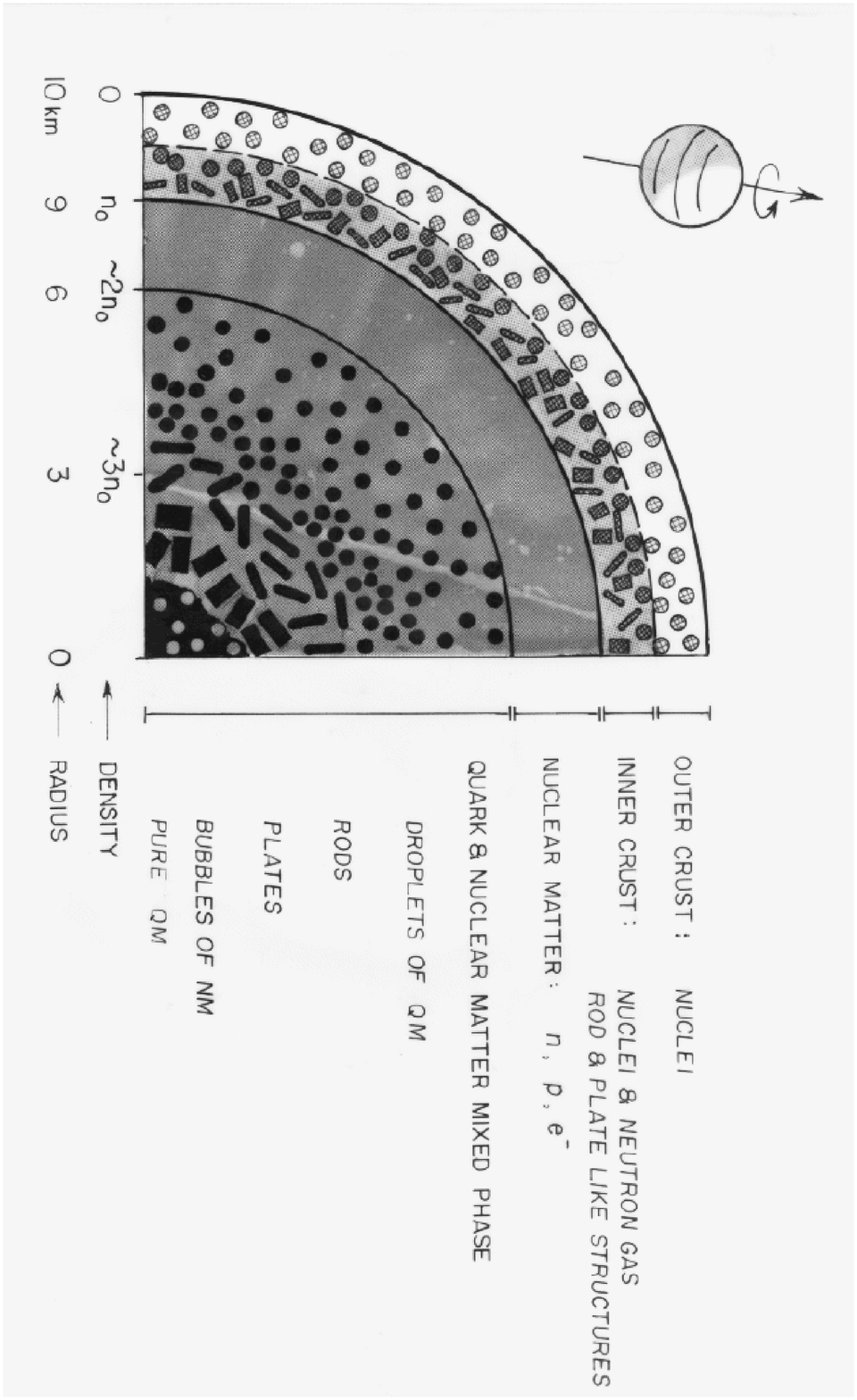,height=85mm,angle=90}}}
\end{minipage}
\vspace*{-5mm}
\begin{center}
\begin{minipage}{13cm}
\baselineskip=12pt
{\begin{small}
Fig.~1. Nuclear and quark matter tructures in a $\sim1.4M_\odot$ neutron star.
Typical sizes of structures are $\sim10^{-14}m$ but have been scaled up
to be seen.
\end{small}}
\end{minipage}
\end{center}

\section{Rotation}
As rotating neutron stars slow down, the pressure and the density in
the core region increase due to the decreasing centrifugal
forces and phase transitions may occur in the center.
At the critical angular velocity $\Omega_0$, where the
phase transitions occur in the center of a neutron star, 
we find that the moment of inertia, angular velocity, braking index
etc.\ change in a characteristic way.

The general relativistic equations for slowly rotating stars were
described by Hartle \cite{Hartle}.  Hartle's equations are quite
elaborate to solve as they consist of six coupled differential
equations as compared to the single TOV equation
in the non-rotating case.  However, Hartle's
equations cannot be used in our case because the first order phase
transition causes discontinuities in densities so that changes are not
small locally. This shows up, for example, in the divergent
thermodynamic derivate $d\rho/dP$.  From Einstein's field equations
for the metric we obtain from the $l=0$ part the generalized rotating
version of the TOV equation \cite{rot}
\begin{equation}
  \frac{1}{\rho+P}\frac{dP}{da} = - G\frac{m+4\pi a^3P}{a^2(1-2Gm/a)} 
                    + \frac{2}{3}\Omega^2 a , 
  \label{PaGR}
\end{equation}
where $m(a)=4\pi\int_0^a\rho(a')a'^2da'$.
In the centrifugal force term
we have ignored frame dragging and other corrections of order
$\Omega^2GM/R\sim 0.1\Omega^2$ for simplicity and 
since they have only minor effects in our case.

The rotating version of the TOV equation
(\ref{PaGR}) can now be solved for rotating neutron stars of a given
mass and equation of state with a phase transition. 
If a first order phase transition occur in the center at $\Omega_0$ 
the moment of inertia is generally \cite{rot}
\begin{equation}
  I(\Omega) = 
  I_0\left( 1+\frac{1}{2}c_1\frac{\Omega^2}{\Omega_0^2} -\frac{2}{3}c_2 
                (1-\frac{\Omega^2}{\Omega_0^2})^{3/2} + ...
      \right) , 
  \label{Igen}
\end{equation}
for $\Omega\le\Omega_0$.
Here, $c_2\sim \Delta\rho\Omega_0^3$ is a small number proportional to the
density difference between the two phases and to the critical angular
velocity to 3rd power. For $\Omega>\Omega_0$ the $c_2$ term vanishes.

The pulsars slow down at a rate given by the loss of rotational energy
which commonly is assumed proportional to the rotational angular
velocity to some power (for dipole radiation $n=3$)
\begin{equation}
  \frac{d}{dt} \left(\frac{1}{2}I(\Omega)\,\Omega^2\right) = 
                  -C \Omega^{n+1}. 
   \label{dE}
\end{equation}
Consequently, as the braking index
depends on the second derivative $I''=dI/d^2\Omega$ of the moment of
inertia and thus diverges as $\Omega$ approaches $\Omega_0$ from below
\begin{eqnarray}
     n(\Omega) &\equiv& \frac{\ddot{\Omega}\Omega}{\dot{\Omega}^2} 
    = n - \frac{3I'\Omega+I''\Omega^2}{2I+I'\Omega}
    \simeq n - 2c_1\frac{\Omega^2}{\Omega_0^2}
    +c_2\frac{\Omega^4/\Omega_0^4}{\sqrt{1-\Omega^2/\Omega_0^2}} \,.\label{n}
\end{eqnarray}
For $\Omega\ge\Omega_0$ the term with $c_2$ is absent.

\section{Cooling}
If the matter undergoes a phase transition at a critical temperature
but is supercooled as the star cools down by neutrino emission, a
large glitch will occur when the matter transforms to its equilibrium
state. If the cooling is continuous the temperature will decrease with
star radius and time and the phase transition boundary will move
inwards.  The two phases could, e.g., be quark-gluon/nuclear matter or
a melted/solid phase. In the latter case the size of the hot (melted)
matter in the core is slowly reduced as the temperature drops freezing
the fluid. Melting temperatures have been estimated in \cite{LPR} for
the crust and in \cite{HPS} for the quark matter mixed phase.
Depending on whether the matter contracts as it freezes as most
terrestrial metals or expands as ice, the cooling will separate the
matter in a liquid core of lower or higher density respectively and a
solid mantle around.  When the very core freezes we have a similar
situation as when the star slows down to the critical angular
velocity, i.e., a first order phase transition occurs right at the
center. Consequently, a similar behavior for the  moment of inertia, angular
velocities, braking index may occur by replacing $\Omega(t)$ with $T(t)$
in Eqs.\ (2-4).

\section{Glitches}
The glitches
observed in the Crab, Vela, and a few other pulsars are probably due
to quakes occurring in solid structures such as the crust, 
superfluid vortices or possibly the quark matter lattice in the
core \cite{HPS}. These glitches are very small $\Delta\Omega/\Omega\sim
10^{-8}$ and have a characteristic healing time.  

In \cite{GPW} a drastic softening of the equation of state by a phase
transition to quark matter leads to a sudden contraction of the
neutron star at a critical angular velocity and shows up in a
backbending moment of inertia as function of frequency. As a result,
the star will become unstable as it slows down, will suddenly decrease
its moment of inertia and create a large glitch.
If the matter supercools and makes a sudden transition to its stable phase
the densities may also change and the star will have to contract or expand.
Consequently, a large glitch will be observed.

\section{Summary}

We have discussed various possible phase transitions in neutron
stars and have argued that we expect several first order order
phase transitions to occur when the topological structure of the
mixed phase change in the inner crust, nuclear and quark matter
mixed phase or Kaon condensates.
If a first order phase transitions is
present at central densities of neutron stars, it will show up in
moment of inertia and consequently also in angular velocities in a
characteristic way.
For example, the braking index diverges as $n(\Omega)\sim
c_2/\sqrt{1-\Omega^2/\Omega_0^2}$.

\end{document}